\documentclass{article}

\pdfoutput=1 

\usepackage[utf8]{inputenc}  
\usepackage[T1]{fontenc}     

\usepackage{arxiv}     
\usepackage{fancyhdr}  
\usepackage{bm}        

\usepackage{soul,color}   
\usepackage{anyfontsize}  

\usepackage{graphicx}    
\usepackage{wrapfig}     
\usepackage{float}       
\usepackage{caption}
\usepackage{subcaption}  

\usepackage{booktabs}        
\usepackage{threeparttable}  

\usepackage{amsmath}    
\usepackage{amsfonts}   
\usepackage{nicefrac}   
\usepackage{microtype}  

\usepackage{hyperref}  
\usepackage{url}       

\usepackage{pgfplots}                
\pgfplotsset{width=10cm,compat=1.9}  
\usepgfplotslibrary{external}
\tikzsetexternalprefix{figures/}     

\title{Data Analysis on Speeding Behavior: The Impact of Auditory Warnings and Demographic Factors}

\author{
  Christian Bank Lauridsen$^{1}$, Mads Greve Andersen$^{1}$, Max-Emil Smith Thorius$^{1}$, Fabricio Batista Narcizo$^{1,2}$ \\
  \\
  $^{1}$Computer Science Department, IT University of Copenhagen, Denmark \\
  \texttt{\{chbl, mgan, maxt, narcizo\}@itu.dk} \\
  \\
  $^{2}$Video Technology, GN Advanced Science, Ballerup, Denmark \\
  \texttt{fbnarcizo@jabra.com}
}

\begin{document}
\maketitle

\begin{abstract}
    %
%
%
%
%
%

Speeding significantly contributes to traffic accidents, posing ongoing risks despite advancements in automotive safety technologies. This study investigates how auditory alerts influence speeding behavior across different demographic groups, focusing on drivers' age and experience levels. Using a mobile application to collect real-time driving data, we conducted a field study in Copenhagen/Denmark that included various driving environments and controlled auditory warnings for speed limit violations. Our results revealed that auditory alerts were unexpectedly associated with an increased frequency and duration of speeding incidents. The impact of these alerts varied by experience level: intermediate drivers showed reduced speeding duration in response to alerts, whereas novice and highly experienced drivers tended to speed for more extended periods after receiving alerts. These findings underscore the potential benefits of adaptive, experience-sensitive alert systems tailored to driver demographics, suggesting that personalized alerts may enhance safety more effectively than standardized approaches.

\end{abstract}

%
%
%
%
%
%

\section{Introduction}
Speeding remains a significant factor in traffic accidents worldwide. An analysis by the Danish Accident Investigation Board in Denmark revealed that excessive speed contributed to 41\% of the 270 traffic accidents analyzed~\cite{HVU2020}. Despite advancements in automotive safety technology, including systems like driving assistance and blind spot detection, speeding poses a severe risk. Although modern technologies have reduced fatality rates by 56\% between 1950 and 2012~\cite{NHTSA2019}, their impact on speed-related accidents remains limited.

Automated safety systems, such as automatic emergency braking and cruise control, have improved vehicle safety by utilizing real-time environmental data. However, current systems must fully address the risks associated with excessive speeding~\cite{Abbas2019}. The Danish Accident Investigation Board further demonstrates that speeding is a key risk factor, particularly for younger drivers, with 32\% accidents involving drivers aged 18 to 19 directly related to speeding.

\subsection{Problem Statement}
This research investigates the factors that influence driver behavior in speeding and assesses whether auditory alerts can effectively reduce the time it takes for a driver to exceed the speed limit. We developed and deployed a mobile application to collect real-time data on vehicle dynamics, such as speed, engine performance, traffic flow, and speed limits. Additionally, the system provides sound alerts to notify drivers when they exceed the speed limit. This research analyzes their speeding patterns and responses to sound alerts by collecting data from various demographic groups. The objective is to identify correlations between driver characteristics (e.g., age, experience) and their reactions to these interventions.

This study addresses the following research questions to explore the potential correlations and causal relationships between driver behavior and the impact of sound alerts.

\textbf{Correlation Research Questions (CO-RQ):}
\begin{itemize}
    \item \textbf{CO-RQ-1:} What is the relationship between the driver's experience level and the duration of speeding incidents?

    \item \textbf{CO-RQ-2:} How does the driver's age correlate with the engine's speed during driving?

    \item \textbf{CO-RQ-3:} Is there a correlation between traffic flow and vehicle speed between different age groups?
\end{itemize}

\textbf{Cause and Effect Research Questions (CE-RQ):}
\begin{itemize}
    \item \textbf{CE-RQ-1:} How does alerting drivers when they exceed the speed limit influence the frequency of speeding incidents?
	
    \item \textbf{CE-RQ-2:} How do speed limit alerts affect the duration of speeding incidents?
	
    \item \textbf{CE-RQ-3:} Does the effect of speed limit alerts on speeding behavior differ among drivers with varying experience levels?
\end{itemize}

These questions guide the study's exploration into the interaction between demographic factors and alert responses, laying the groundwork for the methodological approach outlined below.
%
%
%
%
%
%

\section{Related Work}
Several studies have previously been conducted to analyze driver behavior and induce safer driving by reducing speeding behavior, including auditory alerts.

In a real-world study, Oka et al.~\cite{Oka2019} investigated how auditory warnings influenced prefrontal cortex activity in response to a tunnel construction event. The participants drove through a tunnel twice, once with an auditory alert and once without. The researchers collected data using a Controller Area Network (CAN) device with a GPS for vehicle position. Also, they collected the brain activity with a Functional Near-Infrared Spectroscopy (fNIRS) device. The study found that the auditory alerts increased prefrontal cortex activation, helping drivers maintain smoother control of the accelerator pedal when encountering the construction. These findings suggest auditory alerts promote safer driving behavior by enabling a more controlled response to unexpected road events. However, while this study reveals that the driver notices and acts upon auditory alerts during unexpected road events, it needs to provide insight into how drivers would respond to such alerts when exceeding the speed limit.

A variety of mitigations towards speeding behavior have been experimented on. One study by Kumar and Kim~\cite{Kumar2005} redesigned the dashboard of automobiles to a new speedometer design to change the speeding behavior of drivers. The dynamic speedometer would visually resemble the red max RPM values on the car's accelerometer. However, it would display the current speed limit in real-time instead. The researchers found that further investigating this HCI design could significantly alter speeding behaviors. This follows the idea of alerting the user when the driver exceeds the speed limit. However, the design is limited as a visual in the dashboard, and an inattentive driver can potentially miss it. Another study by Chen et al.~\cite{Chen2024} investigated the impact of speeding behavior when implementing an interactive-affective system. The theory behind this system was to mitigate aggressive driving behavior by letting the driver express their feelings to the emotional mediator. This study found that the device effectively reduced aggressive driver behavior but would require further research.

Studies have also examined how personality characteristics and demographics influence speeding behavior. A survey by Luu et al.~\cite{VanLe2023} measured how demographics and personality affect the speeding behavior of motorcyclists in Vietnam aged 15 to 25. Their results indicated that people aged 18 to 25 were more likely to speed than those aged 15 to 17. Similarly, Kim et al.~\cite{Kim2016} analyzed the driving characteristics of different age groups, specifically young (aged 22-29) and middle-aged (aged 50-57) drivers. The researchers conducted this study in a real-world setting where the test participants drove in the exact type of vehicle equipped with a device that collected data from a GPS, camera, and sensors. The data from the device recorded basic driving information such as speed, RPM, and brakes. The researchers concluded that all the test drives followed the same path on the test road. The results revealed that younger people were significantly more likely to exceed the speed limits than middle-aged people.

The analysis of demographic patterns in speeding behavior is well documented. However, the impact of auditory alerts in conjunction with different contexts, such as demographic groups, has yet to be thoroughly explored.
%
%
%
%
%
%

\section{Methodology}

\subsection{Data Collection Methodology}
We designed a data collection process to gather quantitative data from a diverse range of demographic groups, focusing on different age groups and levels of driving experience. A prior study on aggressive driving behavior, which demonstrated the influence of demographic factors on driving patterns, inspired this approach~\cite{Aljagoub2023}. The experiment included participants ranging from 17 to 80 years old to investigate the correlation between driver age and the engine usage, as outlined in CO-RQ-2. We also included experienced and less experienced drivers, as defined by the research questions CO-RQ-1 and CE-RQ-3.

\begin{figure}[!ht]
    \centering
    \resizebox{0.5\textwidth}{!}{%
        \input{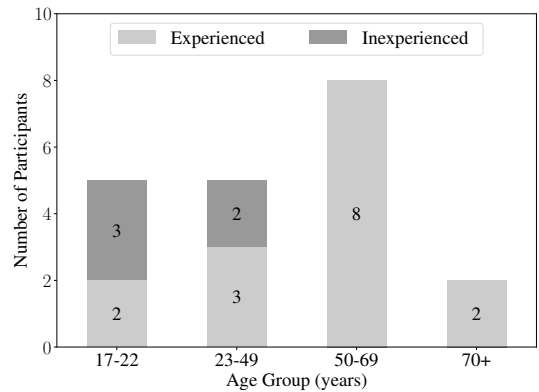}
    }
    \caption{Distribution of participants by age group and driving experience level. The bar chart categorizes participants into four age groups (17-22, 23-49, 50-69, and 70+) and further segments them by their experience level, with ``Experienced'' drivers shown in light gray and ``Inexperienced'' drivers in dark gray. The $X$-axis denotes the age groups, while the $Y$-axis represents the number of participants. Numbers within each bar indicate the count of participants per experience level in each age group. The chart highlights more experienced drivers in the 50-69 age group, with fewer participants in both experience categories for the youngest and oldest age groups. This distribution provides insights into the demographic composition of the study sample, which could influence driving behavior patterns across age and experience levels.}
    \label{fig:stacked_bar}
\end{figure}

Figure~\ref{fig:stacked_bar} illustrates the distribution of participants by age, with the $X$-axis representing age groups and the $Y$-axis indicating the frequency within each group and the participant drive experience. The mean participant age was $41 \pm 19.8$, representing a broad variance of participants' age. However, the study underrepresented certain demographic groups, particularly individuals aged 30 to 49 and 62 to 75. This gap limited the completeness of the dataset. Additionally, there was a higher representation of individuals in their twenties and fifties.

The inexperienced drivers have three or fewer years of experience, and the experienced drivers have four or more years of experience. This demographic variation allowed us to explore the potential differences in behavior across age groups and driving experiences, specifically focusing on how these variables influenced speeding incidents and responses to auditory alerts.

\subsection{Application Setup}
We developed a mobile application, \textit{Safer Driving}, to collect real-time driving data and alert drivers when they exceed speed limits. The application continually saved snapshots of critical variables, including vehicle speed, engine speed, traffic speed, and the speed limit. The source code is publicly accessible on GitHub\footnote{https://github.com/2rius/safer-driving}.

The application's user interface was minimal, as the primary focus was the data collection. The application did not require direct user interaction beyond the initial setup. Upon launch, the experiment conductor inputted key participant details, including the driver's age, occupation, driving experience, and residence. Different participants provided the data, and the system played auditory alerts whenever the vehicle exceeded the speed limit.

\subsection{Auditory Warning System}
The application featured an auditory warning system that issued a single alert when the driver exceeded the speed limit. The warning system activates once per speed violation, avoiding repetitive alerts. This feature raises the driver's awareness without causing distraction. We used the data collected during these events to evaluate the impact of alerts on driving behavior.

\subsection{Driving Protocols}
\begin{figure}[!ht]
    \centering
    \begin{subfigure}[b]{0.375\textwidth}
        \centering
        \caption{}
        \includegraphics[height=0.3\textheight]{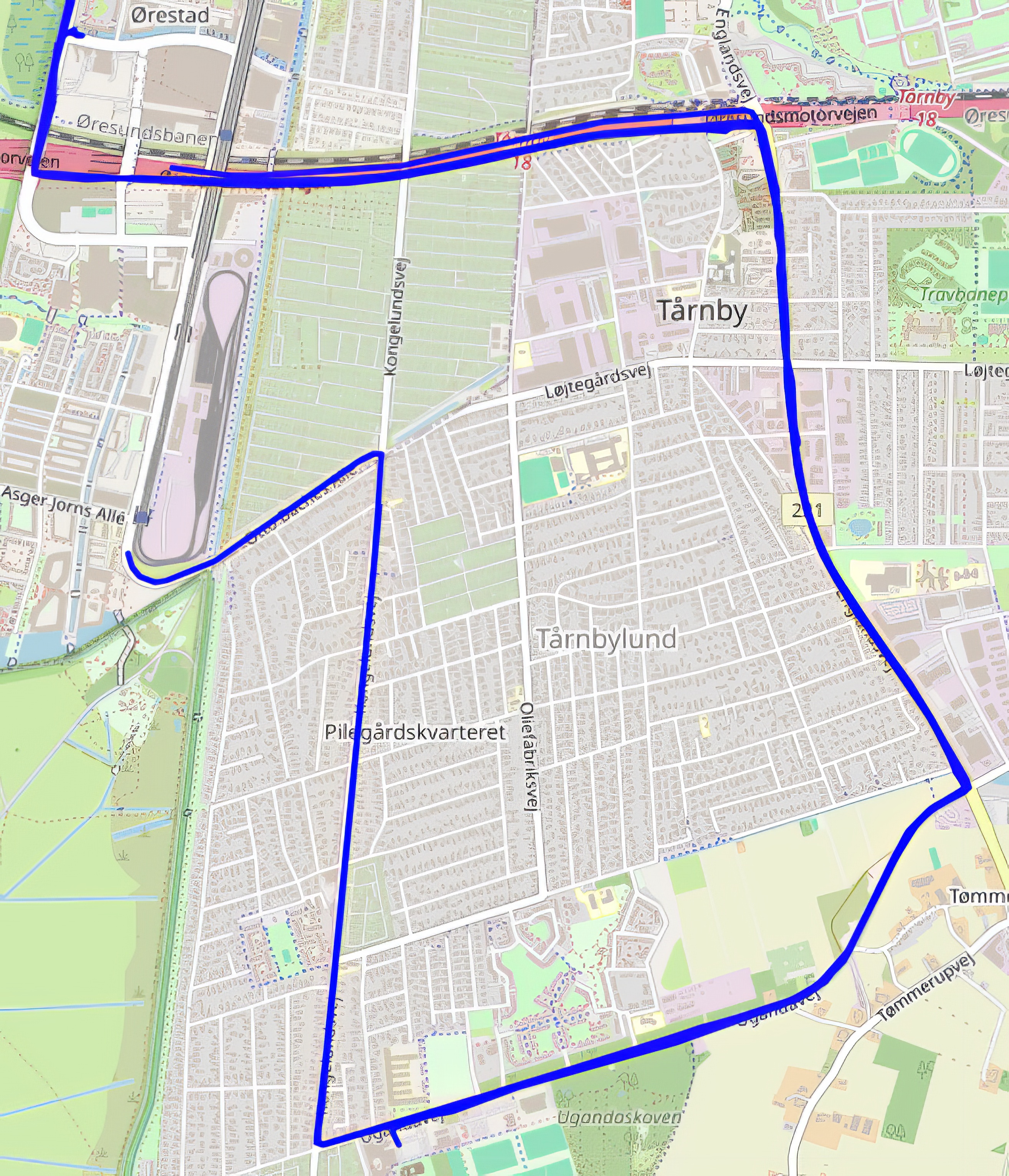}
        \label{fig:route-main}
    \end{subfigure}
    \begin{subfigure}[b]{0.4\textwidth}
        \centering
        \caption{}
        \includegraphics[height=0.3\textheight]{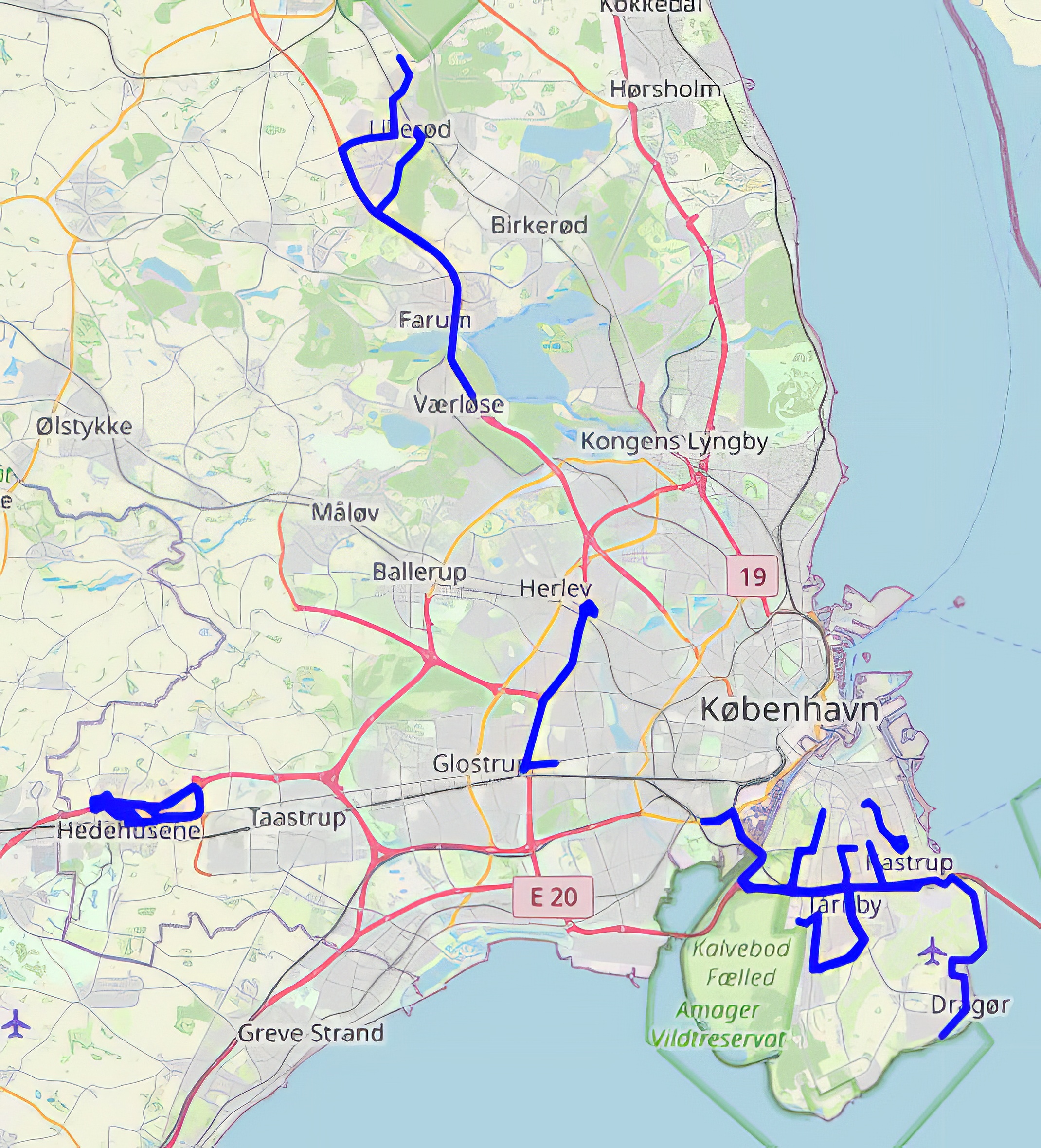}
        \label{fig:route-overview}
    \end{subfigure}
    \caption{The participants drove on the blue routes around Copenhagen and nearby regions as part of the project's data collection process. These routes encompass a mix of environments, including motorways, rural areas, and urban streets, allowing for a comprehensive assessment of driving behavior across different settings. The maps illustrate specific segments of the routes, with each segment chosen to capture varying driving conditions and traffic patterns. \textbf{(a)} This image highlights the primary route around the Copenhagen S and Tårnby areas. \textbf{(b)} This image presents an overview of all routes used in the project, extending beyond Copenhagen into suburban and rural areas.}
    \label{fig:routes}
\end{figure}

Drivers followed predetermined routes to ensure consistent data collection. The routes incorporated different road types---motorway, rural, and city---to compare driving behaviors in various environments. Figure~\ref{fig:route-main} shows an example route followed by five participants. When one route was missing certain road types, additional routes outside Copenhagen (Denmark) supplemented the data, ensuring comprehensive coverage across all road types, as illustrated in blue in Figure~\ref{fig:route-overview}.

\subsubsection{User Research Methodology}
We used the Wizard of Oz paradigm to simulate the experience of driving with an automated speeding alert system. While the participants were unaware of the experiment's details, a ``wizard'' observed and controlled certain elements, such as the audio alerts triggered when the driver exceeded the speed limit. The test environment closely mimics real-world conditions while allowing for controlled manipulation of the audio alert variable. The system reported live data every 2-3 seconds during each session and stored it in a database for later analysis.

\subsubsection{Acquired Variables}
The quantitative data collection focused on three datasets: \textit{live data}, \textit{speeding incidents}, and \textit{ride info}. The \textit{live data} dataset included OBD-II values such as vehicle speed (km/h), engine speed (RPM), and engine load (\%). In contrast, the \textit{speeding incidents} dataset recorded all speeding events. The \textit{ride info} dataset summarized vital information for each ride, including driving time, speeding incidents, and the type of road.
%
%
%
%
%
%

\section{Data Analysis}
This section presents the statistical methods and findings for analyzing the factors influencing driver behavior concerning speeding and assessing the effectiveness of auditory alerts. We analyzed this data to answer the correlation-based and cause-effect research questions below. The Jupyter Notebook for this data analysis is publicly accessible on GitHub\footnote{https://github.com/fabricionarcizo/safer-driving-paper}.

\subsection{Correlation Analysis}
\begin{figure}[!ht]
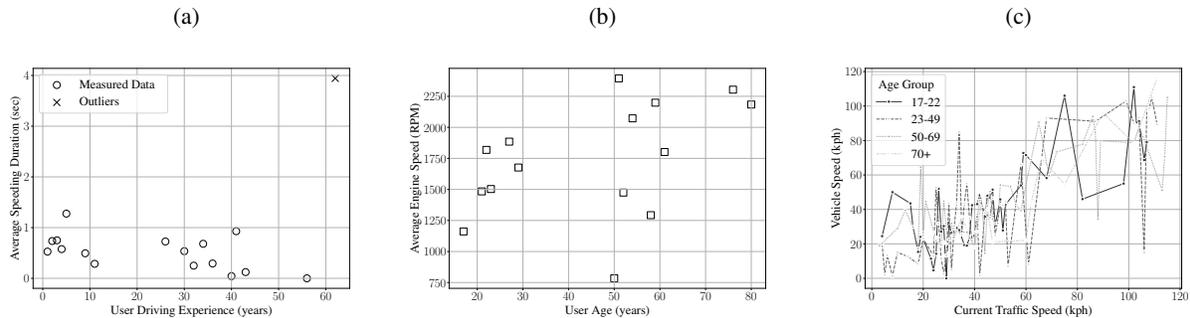

    \centering
    \begin{subfigure}[b]{0.33\textwidth}
        \caption{}
        \label{fig:correlation-analysis-a}
        \resizebox{\textwidth}{!}{%
            \input{pgf/co-rq-1.pgf}
        }
    \end{subfigure}
    \begin{subfigure}[b]{0.33\textwidth}
        \caption{}
        \label{fig:correlation-analysis-b}
        \resizebox{\textwidth}{!}{%
            \input{pgf/co-rq-2.pgf}
        }
    \end{subfigure}
    \begin{subfigure}[b]{0.33\textwidth}
        \caption{}
        \label{fig:correlation-analysis-c}
        \resizebox{\textwidth}{!}{%
            \input{pgf/co-rq-3.pgf}
        }
    \end{subfigure}
    \caption{This set of images illustrates the relationships between driver characteristics (experience and age) and various driving parameters, including speeding behavior, engine speed, and vehicle speed across age groups. These analyses offer insights into how experience and traffic flow impact driving dynamics, potentially influencing safety and efficiency. \textbf{(a)} Scatter plot showing the relationship between user driving experience and the average speeding duration in seconds. \textbf{(b)} Scatter plot depicting the correlation between driver age and average engine speed (RPM) during driving. \textbf{(c)} Line plot showing the relationship between current traffic speed and vehicle speed, segmented by age groups (17-22, 23-49, 50-69, and 70+).}
    \label{fig:correlation-analysis}
\end{figure}

We conducted correlation analyses to investigate the relationships between driver characteristics and driving behaviors. Figure~\ref{fig:correlation-analysis} provides insights into how driving experience, age, and traffic conditions interact with driver behavior.

\subsubsection{CO-RQ-1: Relationship between Driver's Experience Level and Duration of Speeding Incidents.}
To evaluate the research question CO-RQ-1, we calculated the average duration of speeding incidents for different levels of driving experience. Figure~\ref{fig:correlation-analysis-a} shows that one participant in the 70+ age group displayed a significantly distinct driving style, consistently exceeding the speed limit and creating an outlier that skewed the results. To maintain the integrity of the analysis and ensure that findings reflect typical driving behaviors within each age group, we excluded this participant from this correlation analysis. We used Pearson's and Spearman's correlation coefficients to measure the correlation between these two continuous variables, which suggest a moderate negative relationship between driving experience and the duration of speeding incidents.

The Pearson correlation coefficient of $-0.54$ indicates a moderate, negative linear relationship between driving experience and speeding duration. This suggests that as a driver's experience level increases, the duration of their speeding incidents tends to decrease linearly. A negative coefficient means that more experience is associated with shorter durations of speeding incidents, reflecting that more experienced drivers are more cautious or less inclined to speed for extended periods.

The Spearman correlation coefficient of $-0.52$ also suggests a moderate, negative relationship. Spearman's correlation is based on ranks and is less sensitive to linearity, indicating that drivers with higher experience ranks tend to have shorter speeding durations, even if the relationship is not strictly linear. The close values for Pearson and Spearman correlations imply that the relationship is approximately linear but may contain some non-linear patterns where experience still generally leads to shorter speeding durations.

These moderate, negative correlations suggest that as driving experience increases, the duration of speeding incidents tends to decrease. This finding implies that more experienced drivers may have shorter speeding durations due to increased caution, risk awareness, or adherence to safe driving practices. The close alignment of Pearson and Spearman values also suggests that this pattern is consistent across the data, even with minor non-linear variations.

\subsubsection{CO-RQ-2: Correlation Between Driver's Age and Engine Speed (RPM) During Driving.}
For CO-RQ-2, we analyzed the correlation between driver age and average engine speed (RPM). Figure~\ref{fig:correlation-analysis-b} shows a scatter plot with the driver's age versus average engine speed. We measured Pearson's and Spearman's correlation coefficients as $0.45$ and $0.49$, respectively, indicating a moderate positive correlation between age and engine speed.

A Pearson correlation coefficient of $0.45$ indicates a moderate, positive linear relationship between the driver's age and average engine speed. This positive correlation suggests that, generally, as drivers age, engine speed also tends to increase. However, this relationship could be more assertive (it is only moderate).

A Spearman correlation coefficient of $0.49$ also shows a moderate positive relationship. However, it measures the monotonic (rank-based) relationship between the driver's age and engine speed. This positive Spearman coefficient suggests that older drivers have higher average engine speeds than younger drivers, even if the relationship is not linear. 

The slight difference between them could imply a few instances where the relationship is not perfectly linear but generally follows the same trend. For example, some drivers may have higher engine speeds at specific age ranges but not others, creating a mild non-linearity. Older drivers may generally have higher engine speeds, but not strictly linearly. Therefore, age alone does not entirely predict engine speed (other factors, such as driving habits, vehicle type, or road conditions, may also play a role).

\subsubsection{CO-RQ-3: Relationship Between Traffic Flow and Vehicle Speed Across Different Age Groups.}
Finally, for CO-RQ-3, we investigated the relationship between traffic flow (measured by current traffic speed) and vehicle speed across different age groups (17-22, 23-49, 50-69, and 70+ years). Figure~\ref{fig:correlation-analysis-c} shows a line plot, stratified by age group, depicting the variations in vehicle speed at different traffic speeds. We performed a one-way ANOVA to assess differences in vehicle speed across age groups, resulting in an $F$-statistic of $63.84$ and a $p$-value of $<0.001$.

The ANOVA results, with a high $F$-statistic and $p<0.001$, indicate that vehicle speeds differ significantly across age groups. These differences suggest that age could be an essential factor in traffic flow, as variations in speed by age group may contribute to changes in how traffic moves in different areas or times of day. Since ANOVA tells us a significant difference, we conducted post hoc tests using Tukey's HSD to identify which specific age groups differ significantly. Table~\ref{tab:tukey_hsd} gives us an insight into whether certain age groups drive substantially faster or slower than others.
\begin{center}
    \begin{threeparttable}[h]
        \caption{Pairwise comparisons of mean vehicle speed differences across age groups, with adjusted $p$-values and 95\% confidence intervals. Significant differences in mean speeds are observed between most age group pairs, as indicated by the ``Reject'' column. A $True$ value signifies that the null hypothesis (no difference in means) is rejected at the $0.05$ significance level, demonstrating statistically significant speed variations among different age demographics. This analysis provides insights into how vehicle speed varies by age, highlighting notable contrasts between groups such as 17-22 vs. 50-69 and 50-69 vs. 70+.}
        \label{tab:tukey_hsd}
        \begin{tabular}{rrrrrrr}
            \toprule
                Group 1\tnote{*} & Group 2\tnote{*} &  Mean Diff\tnote{\textdagger} &  $p$-Adjusted\tnote{\textdaggerdbl} &    Lower\tnote{\S} &   Upper\tnote{\S} & Reject\tnote{\#} \\
            \midrule
                17-22 &   23-49 &    -3.6712 &        9.7e-3 &  -6.6919 & -0.6504 &   True \\
                17-22 &   50-69 &    10.7488 &             0 &   7.8919 & 13.6057 &   True \\
                17-22 &     70+ &     2.5596 &        3.7e-1 &  -1.5146 &  6.6338 &  False \\
                23-49 &   50-69 &      14.42 &             0 &  11.5943 & 17.2457 &   True \\
                23-49 &     70+ &     6.2308 &        5.0e-4 &   2.1784 & 10.2831 &   True \\
                50-69 &     70+ &    -8.1892 &             0 & -12.1209 & -4.2575 &   True \\
            \bottomrule
        \end{tabular}
        \begin{tablenotes}
            \footnotesize
            \item[*] These two age groups are compared.
            \item[\textdagger] The difference in mean vehicle speed between the two age groups.
            \item[\textdaggerdbl] $p$-value for each pairwise comparison. If this value is below $0.05$, the difference in means is statistically significant.
            \item[\S] These values represent the $95\%$ confidence interval for the mean difference, indicating the range within which the true difference likely falls.
            \item[$\!$\#] A $True$ value means the null hypothesis (that there is no difference between the two groups) is rejected, indicating a statistically significant difference.
        \end{tablenotes}
    \end{threeparttable}
\end{center}

\begin{figure}[!ht]
    \centering
    \resizebox{0.5\textwidth}{!}{%
        \input{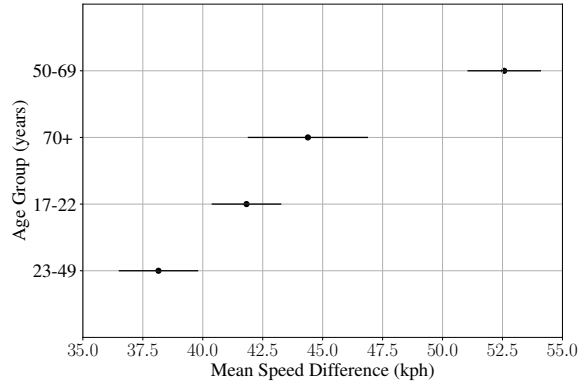}
    }
    \caption{Tukey HSD (Honestly Significant Difference) Test results examine the speed differences across different age groups in driving behavior. The plot shows the mean speed difference for each age group with corresponding confidence intervals, indicating the variance in speed compliance among different demographics. The $X$-axis shows the mean speed difference, with error bars representing the confidence intervals of $95\%$. The $Y$-axis represents the age groups, including categories 17-22, 23-49, 50-69, and 70+. This analysis helps determine if there are statistically significant differences in average speeds between age groups, with visible distinctions, especially between the younger and older age ranges. The intervals that do not overlap suggest significant differences, providing insights into age-related tendencies in driving speeds and speed compliance.}
    \label{fig:tukey-hsd}
\end{figure}

The Tukey's HSD test revealed significant differences in vehicle speed across age groups, with older drivers (50-69) driving the fastest, middle-aged drivers (23-49) the slowest, and both young (17-22) and senior (70+) drivers maintaining moderate speeds, suggesting that age-related variations in speed could influence traffic flow dynamics, as shown in Figure~\ref{fig:tukey-hsd}. The significant variance in the group 70+ is due to the different drive styles between the two participants (one drove too fast and the other too slow). These differences in average speeds across age groups could influence overall traffic flow, as differing speeds among age groups can lead to variations in how smoothly traffic moves.

\subsection{Cause and Effect Analysis}
\begin{figure}[!ht]
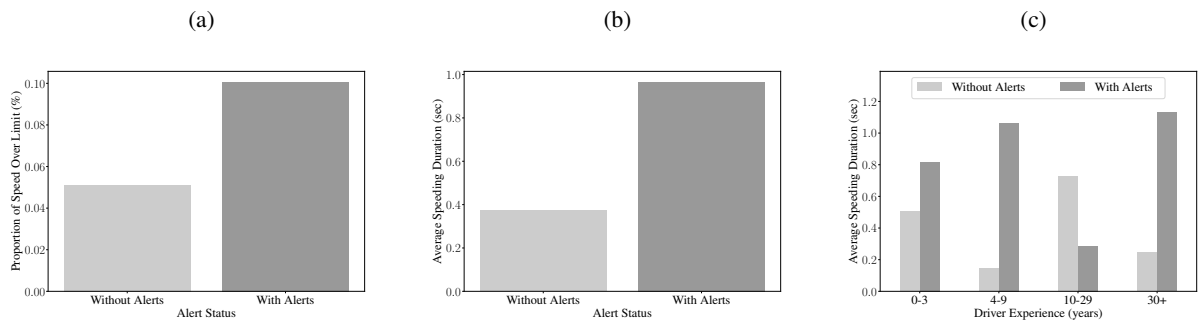

    \centering
    \begin{subfigure}[b]{0.33\textwidth}
        \caption{}
        \label{fig:cause-effect-analysis-a}
        \resizebox{\textwidth}{!}{%
            \input{pgf/ce-rq-1.pgf}
        }
    \end{subfigure}
    \begin{subfigure}[b]{0.33\textwidth}
        \caption{}
        \label{fig:cause-effect-analysis-b}
        \resizebox{\textwidth}{!}{%
            \input{pgf/ce-rq-2.pgf}
        }
    \end{subfigure}
    \begin{subfigure}[b]{0.33\textwidth}
        \caption{}
        \label{fig:cause-effect-analysis-c}
        \resizebox{\textwidth}{!}{%
            \input{pgf/ce-rq-3.pgf}
        }
    \end{subfigure}
    \caption{This series of images evaluates the impact of auditory alerts on speeding behavior, explicitly examining how alerts influence the frequency and duration of speeding incidents and how drivers with varying experience levels respond to alerts. These visualizations provide insights into the effectiveness of speed limit alerts in promoting safer driving behaviors. \textbf{(a)} Bar chart illustrating the influence of alerts on the average frequency of speeding incidents. \textbf{(b)} Bar chart displaying the effect of speed limit alerts on the average duration of speeding incidents in seconds. \textbf{(c)} Bar chart showing the effect of alerts on speeding behavior among drivers with different experience levels.}
    \label{fig:cause-effect-analysis}
\end{figure}

We conducted cause-and-effect analyses to assess the effects of auditory alerts on speeding behavior. Figure~\ref{fig:cause-effect-analysis} shows the impact of speed limit alerts on driver behavior, specifically regarding speeding frequency and duration and how driver experience moderates the effects of alerts.

\subsubsection{CE-RQ-1: Influence of Alerts on the Frequency of Speeding Incidents.}
To evaluate CE-RQ-1, we compared the frequency of speeding incidents with and without alerts. We performed an independent $t$-test due to unequal sample sizes between the groups. The results showed a statistically significant difference, with a $t$-statistic of $6.60$ and a $p$-value of $<0.001$.

A $t$-statistic of $6.60$ suggests a substantial difference in the means of the two groups (drivers with alerts vs. drivers without alerts), as shown in Figure~\ref{fig:cause-effect-analysis-a}. This high $t$-value indicates that the effect size (the influence of alerts on speeding incidents) is relatively large, showing a clear difference in behavior between the two groups.

The $p<0.001$ indicates that the observed difference is doubtful due to random chance, and we can confidently reject the null hypothesis. This means there is a statistically significant difference in the frequency of speeding incidents based on whether drivers receive sound alerts.

These values indicate that auditory alerts significantly influence driver behavior by increasing the frequency of speeding incidents. The substantial $t$-statistic, combined with a very low $p$-value, suggests that the alerts are ineffective in reducing speeding, as drivers who receive alerts speed more frequently than those who do not.

\subsubsection{CE-RQ-2: Effect of Speed Limit Alerts on Duration of Speeding Incidents.}
For CE-RQ-2, we analyzed the duration of speeding incidents with and without alerts. An independent $t$-test revealed a significant difference in speeding duration, with a $t$-statistic of $5.71$ and a $p$-value of $<0.001$. The mean duration of speeding incidents with alerts is $0.966\pm2.5$ seconds, compared to $0.373\pm4.5$ seconds without alerts.

The $t$-statistic of $5.71$ reflects a substantial difference in means, suggesting that alerts significantly impact the duration of speeding incidents. This high value suggests a strong separation between the groups regarding their average speeding duration. The $p<0.001$ indicates that the observed difference is statistically significant and not due to random chance, allowing us to reject the null hypothesis confidently.

These findings suggest that drivers receiving alerts tend to have longer durations of speeding incidents than those without alerts (as shown in Figure~\ref{fig:cause-effect-analysis-b}), contrary to our initial expectations. This could imply that the alerts are prompting drivers to remain at higher speeds for longer or that alerts may need to prompt drivers to reduce their speed immediately and effectively.

\subsubsection{CE-RQ-3: Effect of Alerts on Speeding Behavior Among Drivers with Different Experience Levels.}
Finally, for CE-RQ-3, we investigated the interaction effect between driver experience level (0-3, 4-9, 10-29, and 30+ years of driving experience) and alert presence on speeding behavior, as shown in Figure~\ref{fig:cause-effect-analysis-c}. Table~\ref{tab:anova-results} presents the results of a two-way ANOVA suggesting that the impact of alerts on speeding duration varied by experience level.
\begin{center}
    \begin{threeparttable}[h]
        \caption{Two-way ANOVA results show the effects of driver experience, alert presence, and their interaction on speeding behavior. Significant results were found for both alert presence ($F = 28.94$, $p < 0.001$) and the interaction between driver experience and alert presence ($F = 4.02, p = 0.007$). However, driver experience alone did not have a statistically significant effect ($F = 0.40, p = 0.753$), indicating that the influence of alerts on speeding behavior may vary depending on the driver's experience level.}
        \label{tab:anova-results}
        \begin{tabular}{lrrrr}
            \toprule
                Factor                               & $F$-statistic & $p$-value \\
            \midrule
                Driver Experience (C(experience))    &      0.162057 &  9.2e-01 \\
                Alert Presence (C(alert))            &     28.327812 &  1.1e-07 \\
                Interaction (C(experience):C(alert)) &      4.857132 &  2.2e-03 \\
            \bottomrule
        \end{tabular}
    \end{threeparttable}
\end{center}

The $F$-statistic of $0.16$ and a $p$-value of $0.922$ suggest that driver experience alone does not significantly affect the duration of speeding incidents. This means there is no substantial difference in speeding duration across experience levels without considering the presence of alerts. On the other hand, the $F$-statistic of $28.32$ with a very low $p$-value of $<0.001$ indicates a significant main effect of alert presence on speeding duration. This result shows that, in general, alerts do affect the duration of speeding incidents across drivers, regardless of their experience level.

The interaction between driver experience and alert presence has an $F$-statistic of $4.86$ and a $p$-value of $0.002$, indicating that the impact of alerts on speeding duration varies significantly depending on driver experience. This interaction effect shows that the influence of alerts on speeding duration is not the same for all experience levels.

These findings indicate a complex interaction between experience level and alert presence. Alerts appear to increase speeding duration for drivers with 0-3 years, 4-9 years, and 30+ years of experience, potentially due to differences in confidence or driving habits. However, drivers with 10-29 years of experience respond to alerts by significantly reducing the duration of their speeding incidents. This interaction suggests that alerts may be equally effective for some drivers but could be tailored to enhance effectiveness across different experience levels.

\subsection{Summary of Findings}
The study examined how demographic factors and auditory alerts affect speeding behavior, yielding several vital insights. The analysis showed a moderate negative correlation between driving experience and the duration of speeding incidents, with Pearson and Spearman correlations of $-0.54$ and $-0.52$, respectively. This indicates that as drivers gain experience, they tend to speed for shorter periods over the speed limit, likely due to increased caution or a stronger adherence to safe driving practices

We observed a moderate positive correlation between driver age and engine speed, with Pearson and Spearman coefficients of $0.45$ and $0.49$. These results suggest that older drivers generally maintain higher average engine speeds than younger drivers. Although primarily linear, this relationship includes minor non-linear aspects, hinting that other factors, such as individual driving habits and road conditions, may also influence this trend. Furthermore, differences in vehicle speeds across age groups were significant, as identified by a one-way ANOVA. Older drivers (aged 50-69) drove the fastest, middle-aged drivers (aged 23-49) maintained the slowest average speeds, while both younger drivers (17-22) and senior drivers (70+) maintained moderate speeds. These age-based variations underscore age as an essential factor in shaping overall traffic flow dynamics.

Auditory alerts had a counterintuitive effect on speeding frequency, influencing cause-and-effect relationships. An independent $t$-test ($t = 6.60, p < 0.001$) revealed that drivers receiving alerts engaged in speeding more frequently than those without, suggesting that the alerts may unintentionally encourage rather than deter speeding. Alerts were also associated with a significant increase in the duration of speeding incidents. The analysis, yielding a $t$-statistic of $5.71$ and a $p < 0.001$, demonstrated that drivers receiving alerts tended to sustain higher speeds for more extended periods, indicating that these alerts do not act as immediate deterrents but may instead reinforce prolonged speeding behavior.

Finally, the study revealed a significant interaction between driver experience and alert presence on speeding duration. A two-way ANOVA ($F = 4.86, p = 0.002$) showed that alerts increased the duration of speeding incidents among drivers with minimal experience (0-3 years) and those with extensive experience (30+ years). However, drivers with 10-29 years of experience reduced their speeding duration in response to alerts, highlighting the potential benefits of an experience-sensitive alert system. Tailoring alerts to account for varying experience levels may enhance their effectiveness in promoting safer driving behavior, thereby contributing to a more adaptive and personalized approach to road safety.
%
%
%
%
%
%

\section{Conclusions}
This study examined how auditory alerts influence driver speeding behavior across different demographics. It evaluated the effectiveness of these alerts in moderating speeding incidents. The findings revealed several key insights:
\begin{itemize}
    \item \textit{Impact of Driver Experience and Demographics on Speeding Behavior}: Age plays a substantial role in shaping speeding behavior. Experienced drivers tend to exhibit shorter speeding durations, suggesting increased caution and adherence to safe driving practices. In contrast, younger or less experienced drivers demonstrate prolonged speeding incidents, potentially due to lower risk awareness. Additionally, age-based differences in driving patterns, such as the higher average engine speeds maintained by older drivers, reflect broader demographic tendencies that can impact traffic flow and safety.

    \item \textit{Effectiveness of Auditory Alerts}: The study reveals a counterintuitive trend. The alerts, rather than reducing speeding frequency and duration, were associated with increased speeding incidents and prolonged speeding behavior. This outcome suggests that auditory warnings, as currently implemented, may lack the immediacy or impact required to serve as effective deterrents, potentially encouraging unintended behavioral responses.

    \item \textit{Moderating Role of Driver Experience on Alert Effectiveness}: The driver experience emerged as a significant moderating factor in the impact of alerts. While novice and highly experienced drivers were more likely to maintain extended speeding durations after receiving alerts, those with intermediate experience (10-29 years) responded by reducing their speeding durations. This variability in responses underscores the potential benefits of tailoring auditory alert systems to accommodate different experience levels, enhancing their effectiveness in promoting safer driving behaviors.
\end{itemize}

In conclusion, the study highlights that while auditory alerts have the potential to modify speeding behavior, their current design may not yield the intended safety benefits. Tailoring alert systems to demographic and experiential differences in drivers could improve their efficacy, thereby contributing to safer driving environments. Future research could explore adaptive alert mechanisms and conduct in-depth studies on how demographic-specific alert designs impact driver behavior.

\bibliographystyle{acm}    
\bibliography{references}  

\begin{thebibliography}{1}

\bibitem{Abbas2019}
{\sc Abbas, M.~K., Jung, L.~T., and Abdulla, R.}
\newblock {An Automated Software-Agents System for Detecting Road Speed Limit Offences}.
\newblock In {\em Proceedings of the 2019 $8^{th}$ International Conference on Software and Computer Applications\/} (New York, NY, USA, Feb. 2019), ICSCA '19, Association for Computing Machinery, pp.~538--543.

\bibitem{Aljagoub2023}
{\sc Aljagoub, D., Ardeshir, F., and Karakurt, A.}
\newblock {The Plague of Aggressive Driving: Definitions, Causes, Severity, Consequences, and Solutions}.
\newblock {\em Open Journal of Safety Science and Technology 13\/} (Sept. 2023), 132--151.

\bibitem{Chen2024}
{\sc Chen, K., Fu, X., and Speed, C.}
\newblock {User-Friendly Interactive Affective System to Leverage Aggressive Driving Behavior}.
\newblock In {\em Proceedings of the 2022 $10^{th}$ International Symposium of Chinese CHI\/} (New York, NY, USA, Feb. 2024), Chinese CHI '22, Association for Computing Machinery, pp.~50--61.

\bibitem{HVU2020}
{\sc {Havarikommissionen for Vejtrafikulykker (HVU)}}.
\newblock {Hvorfor Sker Trafikulykkerne? Tværanalyse af 270 Ulykker}, Dec. 2020.

\bibitem{Kim2016}
{\sc Kim, H.~S., Yoon, D.~S., Shin, H.~S., and Park, C.~H.}
\newblock {Driving Characteristics Analysis of Young and Middle-Aged Drivers}.
\newblock In {\em Proceedings of the 2016 International Conference on Information and Communication Technology Convergence (\/} (Washington, DC, USA, Oct. 2016), ISCC '16, IEEE Computer Society, pp.~864--867.

\bibitem{Kumar2005}
{\sc Kumar, M., and Kim, T.}
\newblock {Dynamic Speedometer: Dashboard Redesign to Discourage Drivers from Speeding}.
\newblock In {\em Proceedings of the CHI '05 Extended Abstracts on Human Factors in Computing Systems\/} (New York, NY, USA, Apr. 2005), CHI EA '05, Association for Computing Machinery, pp.~1573--1576.

\bibitem{NHTSA2019}
{\sc {National Highway Trafic Safety Administration (NHTSA)}}.
\newblock {How Vehicle Safety Has Improved Over the Decades}, 2019.

\bibitem{Oka2019}
{\sc Oka, N., Sugimachi, T., Yamamoto, K., Yazawa, H., Takahashi, H., Gwak, J., and Suda, Y.}
\newblock {Impact of Auditory Alert on Driving Behavior and Prefrontal Cortex Response in a Tunnel: An Actual Car Driving Study}.
\newblock In {\em Advances in Neuroergonomics and Cognitive Engineering: Proceedings of the AHFE 2019 International Conference on Neuroergonomics and Cognitive Engineering, and the AHFE International Conference on Industrial Cognitive Ergonomics and Engineering Psychology\/} (Heidelberg, Germany, July 2019), AHFE '19, Springer International Publishing, pp.~373--382.

\bibitem{VanLe2023}
{\sc Van~Le, L., Chu, M.~C., Nguyen, L.~X., and An, N.~M.}
\newblock {Differences in Personality Characteristics, Demographics, and the Predictive Value of the Self-Reported Speeding Behavior Model of Young Riders in Vietnam}.
\newblock In {\em Proceedings of the 2024 $3^{rd}$ International Conference on Sustainable Civil Engineering and Architecture\/} (Heidelberg, Germany, Dec. 2023), ICSCEA '23, Springer International Publishing, pp.~1757--1764.

\end{thebibliography}

\end{document}